# Volume analysis of supercooled water under high pressure

Solomon F. Duki[1] and Mesfin Tsige[2]

[1]National Center for Biotechnology Information, National Library of Medicine and National Institute of Health, Bethesda MD, 20894 USA

[2]Department of Polymer Science, The University of Akron, Akron OH, 44325 USA

Motivated by an experimental finding on the density of supercooled water at high pressure [O. Mishima, J. Chem. Phys. 133, 144503 (2010)] we performed atomistic molecular dynamics simulations study of bulk water in the isothermal-isobaric ensemble. Cooling and heating cycles at different isobars and isothermal compression at different temperatures are performed on the water sample with pressures that range from 0 to 1.0 GPa. The cooling simulations are done at temperatures that range from 40 K to 380 K using two different cooling rates, 10 K/ns and 10 K/5 ns. For the heating simulations we used the slowest heating rate (10 K/5 ns) by applying the same range of isobars. Our analysis of the variation of the volume of the bulk water sample with temperature at different pressures from both isobaric cooling/heating and isothermal compression cycles indicates a concave-downward curvature at high pressures that is consistent with the experiment for emulsified water. In particular, a strong concave down curvature is observed between the temperatures 180 K and 220 K. Below the glass transition temperature, which is around 180 K at 1GPa, the volume turns to concave upward curvature. No crystallization of the supercooled liquid state was observed below 180 K even after running the system for an additional microsecond.

**I-INTRODUCTION**

Water is a very important substance that exists naturally in its different phases in a wide range of temperature and pressure. This made it the subject of intense study in many interdisciplinary research areas that range from biological systems to large scale industrial application materials. One of the many questions that has been investigated over the years by both theorists and experimentalists is the understanding of the completely different behavior of water observed at low temperature compared to other liquids under the same set of conditions [1-4]. In particular, the structural change in supercooled water under high pressure has been examined from two different point of views that debate whether the liquid-liquid transition at low temperature is continuous or discontinuous. The discontinuity hypothesis asserts the existence of a liquid-liquid critical point (LLCP)[5] where transition between two liquids is discontinuous at the critical point. On the other hand, the singularity-free theory (SF) hypothesized that the liquid-liquid transition is continuous and singularity-free[6]. A recent experimental work on emulsified water by O. Mishima[7] showed that the temperature dependence of the volume of a supercooled water at different pressures has a concave-downward curvature above $\approx$ 200 MPa supporting the LLCP theory.

Recently, Abascal and Vega[8] have compared simulation results with Mishima's experimental work and concluded that the simulation results using the TIP4P/2005 water model supports the existence of a critical point in the supercooled region for real water. However, their simulation was limited to intermediate and high temperatures and did not directly address whether the temperature dependence of the volume of their sample of water is concave-downward at high pressures as observed in the experimental work. Motivated by this open question, we have done molecular dynamics simulations on bulk water sample to examine the phase behavior of water under various temperature and pressure conditions. Our simulation uses both a 3 point and a 4-point water models, discussed below, to model bulk water in isobaric cooling/heating and isothermal compression cycles.

We have organized the manuscript as follows: In Sec. II we outline our methodology and details of the numerical simulations. The resulting simulations data for the cooling, heating and isothermal compression procedures are presented and discussed in sections III, IV and V, respectively. To investigate the structure of the amorphous ice water, we have calculated the radial distribution functions for both the isobaric and isothermal cycles and are reported in Sec. VI. Summary and conclusion are given in Sec. VII.

**II-SIMULATIONS**

We performed all-atomistic Molecular Dynamics (MD) simulations for a sample of 1500 water molecule using LAMMPS[9] in the isothermal-isobaric (NPT) ensemble. The sample we considered is large enough to represent bulk water sample and generates statistically reliable information for our study. In fact, the volume of the initial sample prepared at high temperature has stabilized to its equilibrium value much faster than a sample with 500 water molecules. This additional computational cost we have incurred due to the large system size is offset by the gain in equilibration time and the better representation of the amorphous state of water.

In order to investigate how the results of this work depend on the modeling approach, we have used two different water models. The first model we used is the simple point charge (SPC/E) model[10], which is a 3-point water model and has been successfully used to simulate different aspects of water including thermodynamics study[11], phase diagram (of amorphous ice)[12] and glass temperature[13]. The second model we used is the TIP4P/2005 water model[14]. This is basically a 4-point rigid water model where the oxygen and hydrogen atoms are the three sites. In this model the fourth site, called the M site, is co-planer with the oxygen and hydrogen sites and is located at a distance $d_{OM}$ from the oxygen atom on the bisector line of the H-O-H angle. Compared to other water models TIP4P/2005 model is well known to give the best agreement with experiment for a wide range of states[15].

All simulations were run with periodic boundary conditions in all directions and the velocity-verlet[16] algorithm was used to integrate the particles' equations of motion with a time step of 1 *fs*. The long range Coulombic interactions were calculated with a particle-particle/particle-mesh algorithm[17], while the van der Waals interactions were cutoff at 10 Å. During the simulations the temperature and pressure were regulated by using a Nosè-Hoover thermostat and barostat[18, 19]. In both models, the initial liquid configuration of the sample was prepared at 380 K (close to boiling point) by running it for at least for 1 *ns*. Then two different cooling rates, namely fast cooling (10 K/ns or $10^{10}$ K/s) and relatively slow cooling (10 K/5 ns or 2 x $10^9$ K/s), were used to cool down the sample at each isobar; this process was repeated for different pressures that range from 0 to 1.0 GPa. In each cooling process thermodynamic data was collected at every 1 *ps* for generating the volume versus temperature curve. Some selected temperatures below the glass temperature were run for more than 300 *ns*.

In the glassy state, a finite size system typically needs a very long simulation time in order to reach its equilibrium state. In fact, one has to observe crystallization process as the system reaches the true equilibrium state. We heated the cooled sample for the TIP4P/2005 water model using the slower heating rate (10 K/5 ns) in order to observe the expected hysteresis between cooling and heating cycles around the glass temperature. For particular case of the maximum applied pressure (1.0 GPa) we also re-cooled the heated sample, starting from 250 K, in order to check the effect of thermal history on the curvature of the specific volume.

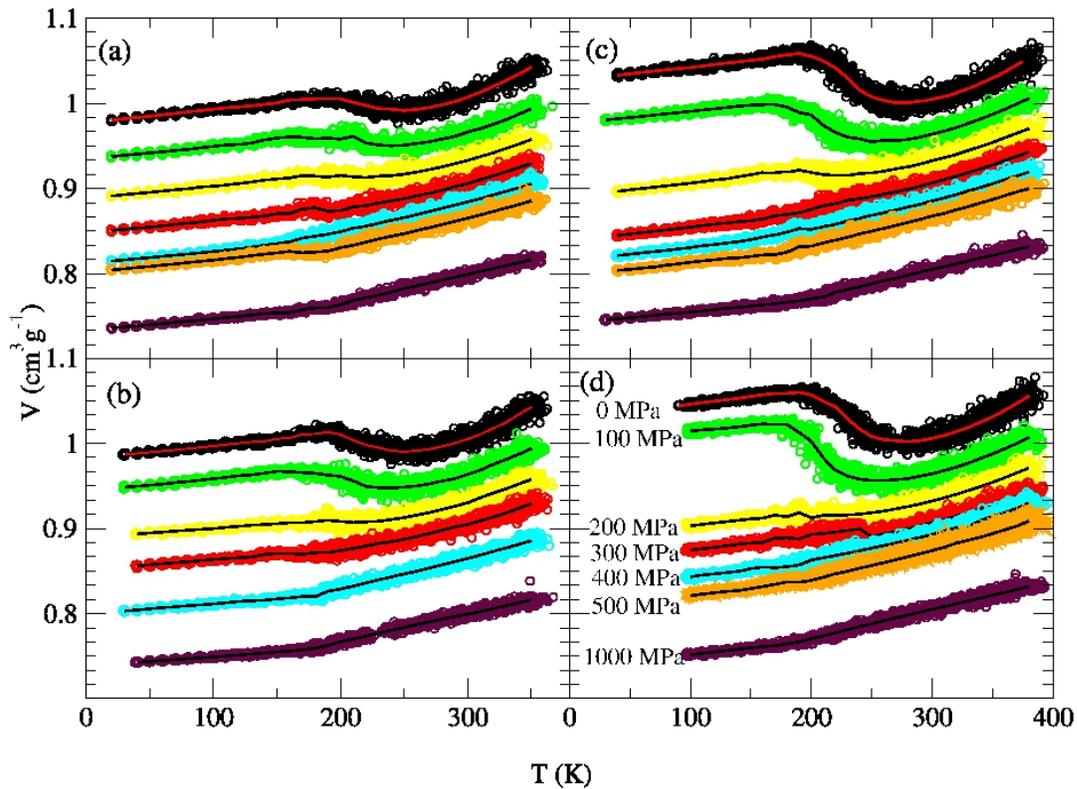

Fig. 1 The P-V-T relation of water from our simulations. The left side of the figure (a and b) shows the results from the SPC/E water model with cooling rates of 10 K/ns and 10 K/5ns, respectively. Similarly, the right side of the figure (c and d) are for TIP4P/2005 water model with cooling rates of 10K/ns and 10K/5 ns, respectively. The different colors show the raw data for the different isobars and the solid line for each isobar is an average of the corresponding raw data. For a given isobar, the same color is used for all the cases.

To investigate how the volume responds to pressure change and if there is a pressure induced phase change in the sample, we have performed isothermal compression simulations at different isotherms on the water sample that was cooled at zero isobar. Simulations ranging from 10 *ns* (at low pressures) to 80 *ns* (at high pressures) were required in order to stabilize the volume at a given pressure. In addition, since we also want to check the curvature of the specific volume generated from this approach, the isotherm temperatures used in this part of the simulation are both below and above the glass temperature. Furthermore, we looked at the compression effect, at 160 K, in inducing a transition from low density amorphous (LDA) to high density amorphous (HDA) at intermediate pressure. In all of the isothermal compression simulations we used the TIP4P/2005 water model by changing the pressure from 0 to 1.0 GPa in increments of 0.1 GPa.

### III- COOLING

The simulation results for the temperature dependence of the volume of the water during the cooling of the sample while it is under various pressure are shown with raw data points in Fig. 1 for the two water models and for the two cooling rates. The different isobars are represented by different colors but for a given isobar the same color is used for all the cases. The solid line for each isobar is an average of the raw data from five independent runs. In averaging the raw data points for a given temperature and pressure, we dropped the first 200 data points to eliminate the initial transient period in the cooling process and used only the last 800 (4800) data points corresponding to the 10 K/ns (10 K/5 ns) cooling rate, respectively. Parts (a) and (b) of Fig. 1 show the specific volume of water from simulations using the SPC/E water model with a cooling rate of 10 K/ns and 10 K/5 ns, respectively. Similarly, parts (c) and (d) show simulations results using the TIP4P/2005 water model with a cooling rate of 10 K/ns and 10 K/5 ns, respectively. Clearly, all the cases in Fig. 1 show the same specific volume of water for its curvature at high pressures independent of the two cooling rates and the two water models used in the current study.

The glass temperature $T_g$ can be inferred from the P-V-T curves shown in Fig. 1 [20], though this method is not straight forward for water at low pressures. Instead the pressure dependent $T_g$ values are determined from isobaric specific heat capacity $C_p(T)$ and mean-squared displacement data as shown in Figs. 2 and 3, respectively, for the TIP4P/2005 water model with the 10K/5 ns cooling rate. $T_g(P)$ values determined from these two approaches are in good agreement and the values are 186 K (0 MPa), 180 K (100 MPa), 177 K (200 MPa), 173 K (300 MPa), 174 K (400 MPa), 178 K (500 MPa), 183 K (1 GPa) with standard deviation ranging from 7 K to 10 K. This anomalous pressure-dependence of water $T_g$ has been recently reported and discussed from results obtained with the SPC/E water model[21]. It is important to note that there is a change in the curvature of the volume curve at these $T_g$ values for all pressures as can be seen in Fig. 1. The calculated $T_g$ values from simulations are generally larger than experimental values mainly due to several orders of magnitude difference in cooling rates between simulations and experiments.

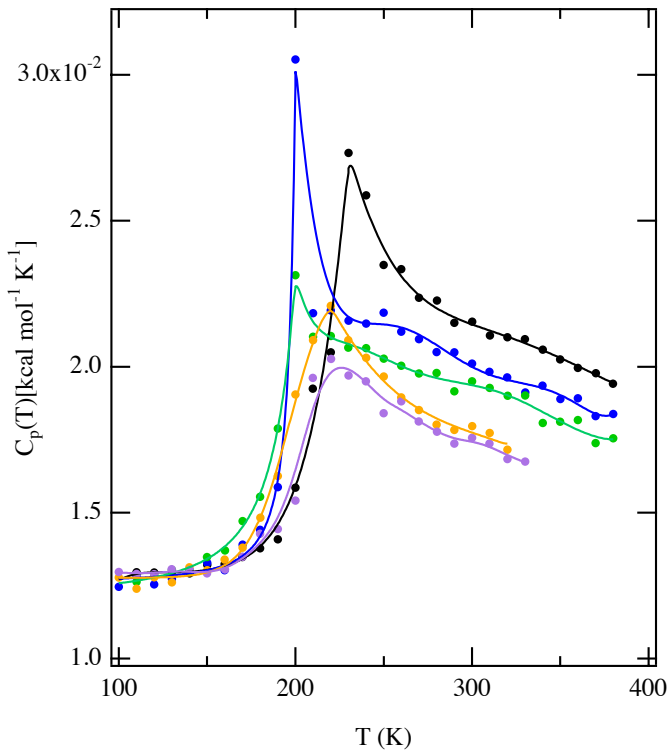

Fig. 2 Temperature dependence of specific heat capacity $C_P$ calculated from the fluctuation of the enthalpy of the system as $c_P=<\delta H^2>/K_B T^2$ at 0 MPa (black), 100 MPa (blue), 200 MPa (green), 500 MPa (orange), and 1 GPa (purple). $T_g(P)$ is defined as the inflection temperature of $C_P(T)$ below its maximum value[13].

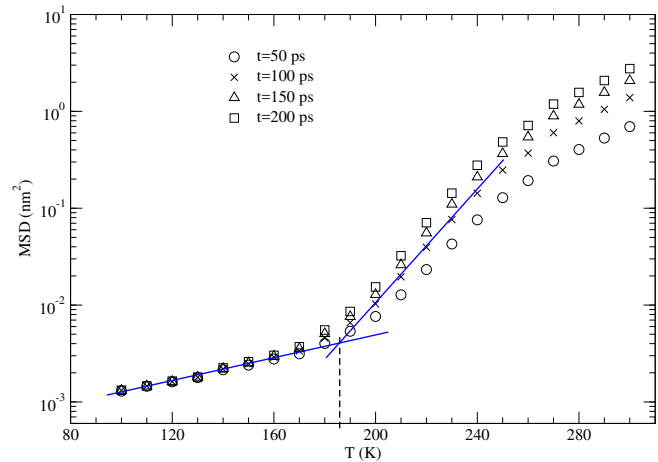

Fig. 3 Mean-squared displacement (MSD) as a function of temperature at four different observation times for 0 MPa case. $T_g$ is estimated from the intersection of the two solid lines and the same protocol was applied for the other pressures to determine $T_g(P)$. The behavior of the mean-squared displacement below $T_g$ is distinctly different from that above $T_g$.

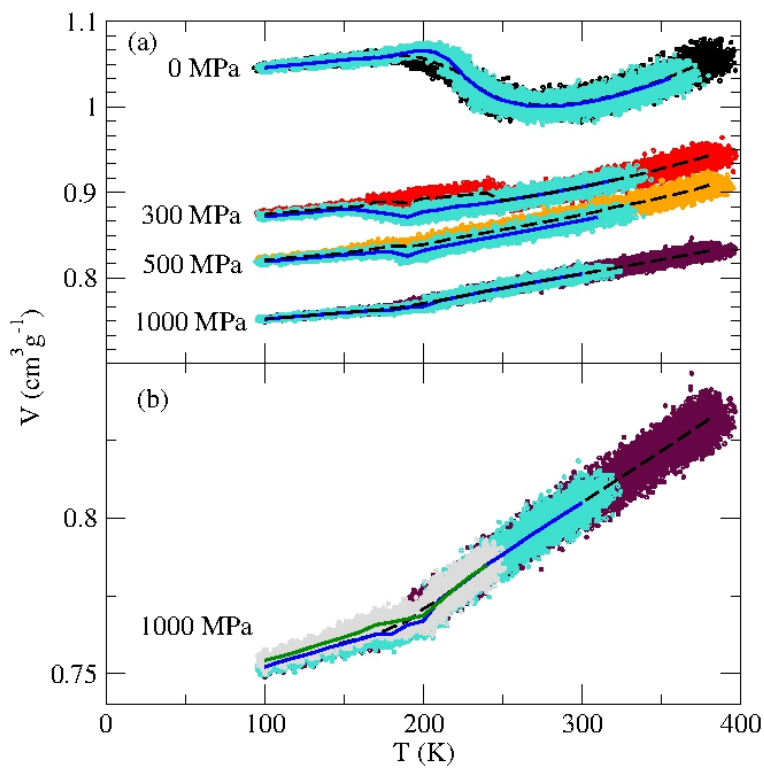

Fig. 4 Selected specific volume versus temperature data for TIP4P/2005 water model during heating/cooling cycles at a rate of 10 K/5 ns. (a) Both the cooling (same color as in Fig. 1 (d)) and heating (shown in cyan color) cycles give the same curvature of the specific volume. The black dotted lines are the averages from the cooling procedures, already shown in Fig. 1 (d), and the solid lines are the averages for the raw data of the heating cycle. (b) Here a specific case of the 1.0 GPa isobar is shown where the cooling is repeated starting from 250 K (shown in gray color) after the sample was heated (shown in cyan color). Clearly the trend for the curvature of the specific volume is reproduced during the re-cooling cycle.

## IV-HEATING

We heated the glass water system to study how the cooled water sample dynamics respond to heating under different isobars. Fig. 4(a) shows simulation data of the specific volume for representative isobars of 0, 0.3, 0.5 and 1.0 GPa. In this figure, for comparison, we have shown both the cooling and heating simulations data together for the TIP4P/2005 water model with a heating/cooling rate of 10 K/5 ns. In the figure, all the heating data are represented by a cyan color while the cooling data are shown by the same colors already used in Fig. 1 (d). The black doted lines are the averages of the volume from the cooling procedures and the blue solid lines are the averages of the volume from the heating procedures. The data from both heating and cooling procedures show the same trend for the curvature of the specific volume at low temperature. There is a hysteresis around the glass temperature for all the isobars studied. Such a hysteresis is expected for simulation of finite size system as it takes an exponentially large amount of simulation time to obtain the true equilibrium state for amorphous ice. In Fig. 4 (b) we have shown a particular case of the 1.0 GPa isobar simulation where the sample that was heated was re-cooled back to 100 K starting from 250 K. In this figure the gray color represents the data from re-cooling procedure and the green solid curve is its average value. Clearly the same trend is followed by the curvature of the specific volume in a consistent manner in all the cooling, heating and re-cooling procedures. A closer look at the data (see Fig. 5) shows strong concave down between 180 K and 220 K, and below the glass transition temperature it turns to concave upward curvature.

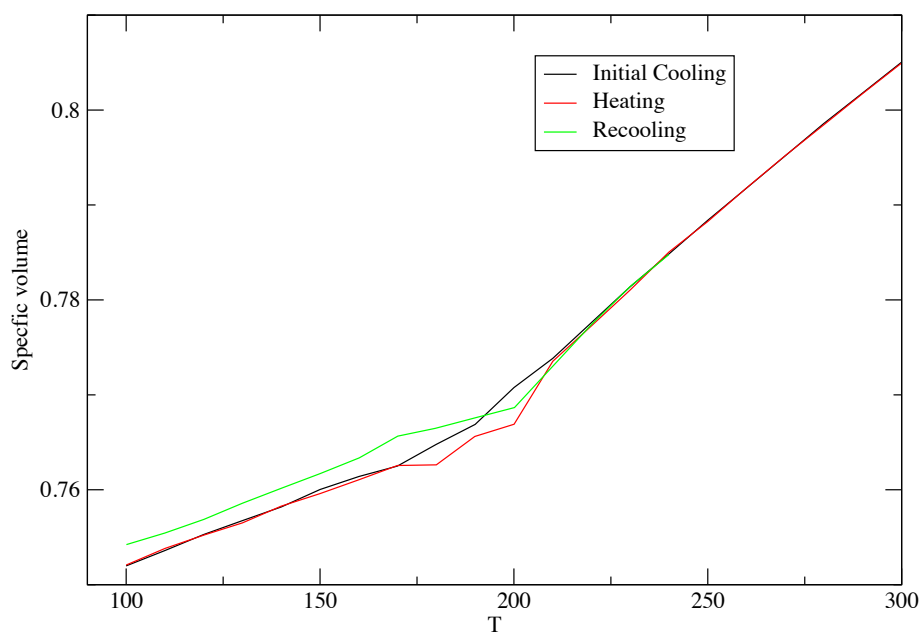

Fig. 5 This is the same figure 1 (b) and a closer look shows strong concave down between 180 K and 220 K. The glass transition temperature is around 180 K at 1 GPa where the specific volume tends to turn to concave upward curvature.

### V- ISOTHERMAL COMPRESSION

Here in this section, we report our investigation of the response of the volume of water during an isothermal compression for different isotherms. In these simulations, the initial water configuration that was equilibrated along the zero isobar during the cooling cycle presented above was set to undergo an isothermal compression at five different temperatures by increasing the pressure in steps of 0.1 GPa. At each pressure and isotherm, we run the system until the volume of the system stabilizes during the timescale of our simulations and this process at each isotherm was repeated until the increased pressure reached 1.0 GPa. Fig. 6 shows the response of the density or the volume of water to increased pressure for five different isotherms. As expected, Fig. 6 (a) shows that at low pressure the density of the glass (at T=160 and 180 K) is smaller than the density of the liquid water (at T=270, 300 and 350 K). The density of the liquid water increases monotonically with increase in pressure. For the amorphous ice the density has a modest change for an increment of 0.1 GPa from its initial zero pressure. However, between 0.1 and 0.2 GPa the density has a sudden jump as reflected in the volume vs temperature plot in Fig. 6 (b). This is a reflection of the sudden structural change taking place in the sample that was induced by pressure transforming it from LDA to HDA, a result consistent with earlier studies[22-24]. Beyond 0.2 GPa, the density of the amorphous ice increases monotonically with increase in pressure. The corresponding structural changes with pressure for the amorphous ice are discussed in more detail in the next section.

Fig. 6 (b) shows the variation of the volume of the sample with pressure where the different colors of the closed circles represent the different isotherms as shown by the same colors in Fig. 6 (a). At a given isotherm, each closed circle represents the volume of the water sample as the pressure is increased from 0, at the top, to 1.0 GPa, at the bottom, (in a vertical direction) in increments of 0.1 GPa. For comparison, the black solid curves are from the isobaric cooling procedure shown in Fig. 1 (d). For a given isotherm, we observe excellent agreement between the data generated by isothermal compression and isobaric cooling.

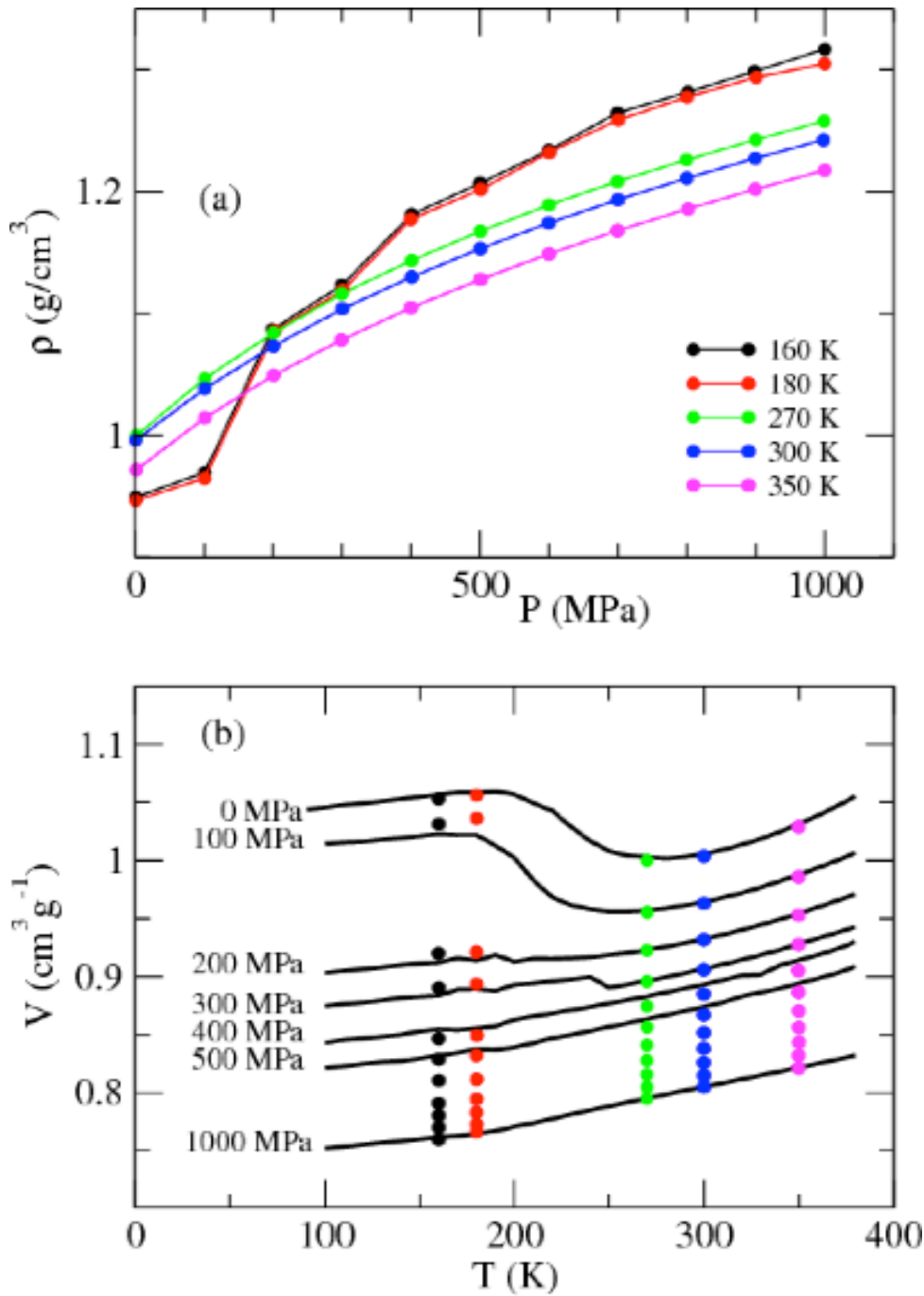

Fig. 6 Density of water as a function pressure during isothermal compression. At low pressure the glassy water has smaller density as expected. At high pressure there is a structural change in the sample where an increase in pressure from 0.1 to 0.2 GPa results a sharp increase in the density. This is a reflection of the pressure induced sudden structural change in the sample where a transition in phase from LDA to HDA is taking place. (b) Specific volume vs temperature during the isothermal compression simulation (shown in color). The different colors of the closed circles show the different isotherms where pressure increases from zero, at the top, to 1.0 GPa, at the bottom, with increment of 0.1 GPa. For comparison, the black solid lines are simulation data from the isobaric cooling shown in Fig. 1 (d). The curvature of the specific volume from the isothermal compression simulation is consistent with the curvature shown from the isobaric cooling procedure.

# VI-RADIAL DISTRIBUTION FUNCTION

To get a better picture of the structural change of the amorphous ice with a change in pressure, we have calculated the radial distribution function (RDF) between the O atoms of the water molecules at 160 K. We performed this calculation for all the cycles: isobaric cooling, isobaric heating and isothermal compression. Fig. 7 shows the RDF at different pressures during isobaric cooling cycle at 160 K. At zero pressure the RDF has its first, second and third maxima located at 2.75, 4.45 and 6.75 Å, respectively. The structure of the RDF does not change when the pressure is increased from 0 to 0.1 GPa. However, the second peak in the RDF drops significantly when the pressure is increased from 0.1 to 0.2 GPa as shown in the figure by the arrow which reflects a change in the second nearest neighbors of the oxygen atoms. Further increase in pressure broadens the second maxima between 3.0 Å and 5.0 Å for different isobars while shifting the peak to the left. As pressure increases further beyond 0.4 GPa, a new maximum starts to develop between 3.0 Å and 3.5 Å where the amplitude of this new peak increases with increasing pressure as indicated by the arrow. This new peak reflects an increase in coordination number between the oxygen atoms resulting a change in the structure of the amorphous ice. The drop in the second peak and the emergence of a new peak in the RDF for higher pressure is consistent with the observed sudden jump of the density during the isothermal compression as shown in Fig. 6 (a).

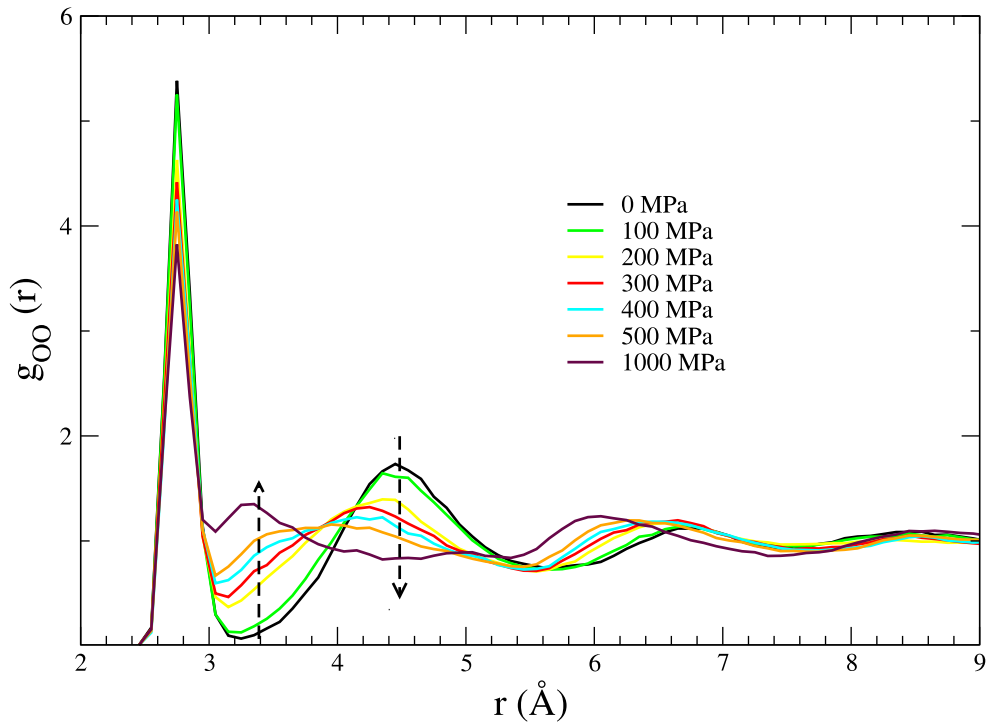

Fig. 7 Radial distribution functions (RDFs) between the O atoms of the water molecules calculated at 160 K during isobaric cooling cycles. The different colors show the different pressures. For increased pressure the first peak decreases and at the same time the second peak starts to shift to the left while broadening its width. Beyond 0.4 GPa a new peak starts to emerge between 3.0 Å and 3.5 Å.

# VII-SUMMARY AND CONCLUSION

In summary, we studied bulk water using atomistic molecular dynamics simulations at isothermal-isobaric ensemble using two different standard water models. We examined the temperature dependence of the volume of the bulk water for different pressures. We have done this first at fixed isobars by cooling the sample starting from high temperature. Then we heated the cooled sample under the same set of isobars to check if the heating and cooling reproduces the same results. The simulation results showed that the specific volume has a concave-downward curvature for supercooled water at high pressure. This result is consistent for both the cooling and heating procedures. We have also studied how the bulk water volume changes under an isothermal compression. Our analysis of the specific volume during the isothermal compression also gives a result consistent with the isobaric cooling/heating results where the curvature at high pressure is concave downward. All in all, the main result of our study does agree with the experimental observation for the curvature of the specific volume of water at high pressure. Specifically our simulation shows strong concave down between 180 K and 220 K. However, below the glass transition temperature it turns to concave upward curvature may be due to the high rate of cooling. We have also observed a pressure induced phase transformation from our examination of the density of water as a function of pressure during isothermal compression. The results from the simulations show that the amorphous ice changes its phase from LDA to HDA as pressure is increased. This phase transformation has been confirmed from our calculation of the Radial distribution function (RDF) of the O atoms of the water molecules at 160 K as a function of pressure for both the isobaric cooling/heating and isothermal compression cycles. The results from the RDF calculations from the different cycles show that significant structural changes in the amorphous ice take place as pressure is increased where the amplitude of the first peak is significantly reduced while the locations of the other peaks are shifted to the left. In particular, the second peak of the RDF observed between 3.0 and 5.0 Å at low pressures drops and shifts to lower $r$ and grows substantially around $r$=3.2 Å. Finally, estimating the LLCP temperature and pressure using the TIP4P/2005 water model is not the focus of the present study and this will be addressed in future investigations.

Acknowledgments

MT's work was supported by ACS Petroleum Research Fund (ACS PRF 54801-ND5).